\documentstyle[aps,epsf,twocolumn]{revtex}
   \begin{document}
   \def\I.#1{\it #1}
   \def\B.#1{{\bbox#1}}
   \def\C.#1{{\cal #1}}
\title{Iterated Conformal Dynamics and Laplacian Growth}
   \author {Felipe Barra, Benny Davidovitch and Itamar Procaccia}
   \address{Department of~~Chemical Physics, The Weizmann Institute of
   Science, Rehovot 76100, Israel}
   \maketitle

%
%
\begin{abstract}
The method of iterated conformal maps for the study of Diffusion
Limited Aggregates (DLA) is generalized to the study of Laplacian
Growth Patterns and related processes. We emphasize the fundamental
difference between these processes: DLA is grown serially with
constant size particles, while Laplacian patterns are grown by
advancing each boundary
point in parallel, proportionally to the gradient of the Laplacian
field. We introduce a 2-parameter family of growth patterns
that interpolates between DLA and a discrete version of
Laplacian growth. The ultraviolet putative finite-time singularities
are regularized here by a minimal tip size, equivalently for 
all the models in this family. With this we stress that the
difference between DLA and Laplacian growth is NOT in the 
manner of ultraviolet regularization, but rather in their
deeply different growth rules. The fractal
dimensions of the asymptotic patterns depend continuously on the
two parameters of the family, giving rise to a ``phase diagram" in which
DLA and discretized Laplacian growth are at the extreme ends. In particular we 
show that the fractal dimension of Laplacian growth patterns is
much higher than the fractal dimension of DLA, with the possibility
of dimension 2 for the former not excluded.
\end{abstract}
\section{Introduction}
This paper had been motivated by an apparent consensus on DLA and 
Laplacian growth Patterns being in the same universality class in terms
of their asymptotic fractal dimensions \cite{84Pat}. We present here a theory
of these processes in two dimensions which clarifies the
differences between these processes, showing in particular that their
asymptotic fractal dimensions differ.

Laplacian Growth Patterns are obtained when the boundary $\Gamma$ of
a 2-dimensional domain is grown
at a rate proportional to the gradient of a Laplacian field $P$.
Outside the domain
$\nabla^2 P=0$, and each point of $\Gamma$ is advanced at a rate
proportional to $\B.\nabla P$ \cite{58ST,84SB}. In Diffusion Limited
Aggregation (DLA) \cite{81WS} a 2-dimensional cluster is grown 
by releasing fixed size random walkers from infinity,
allowing them to walk around until they hit any particle belonging to
the cluster. Since the particles are released one by one
and may take arbitrarily long time to hit the cluster, the probability
field is stationary and in the
complement of the cluster we have again $\nabla^2 P=0$.
The boundary condition at infinity is the same for the
two problems; in radial geometry as $r\to \infty$ the flux
is $\B.\nabla P={\rm const}\times\hat r/r$. Since the probability
for a random walker to hit the boundary is again proportional to 
$\B.\nabla P$, one could think that in the asymptotic limit when the
size of the particle is much smaller than the radius of the cluster,
repeated growth events lead to a growth process which is
similar to Laplacian Growth. Of course,
the ultraviolet regularizations in the two processes were taken
different; in studying
Laplacian Growth one usually  
solves the problem with the boundary condition $P=\sigma\kappa$
where $\sigma$ is the surface tension and $\kappa$ the local
curvature of $\Gamma$ \cite{86BKT}. Without this 
(or some other) ultraviolet regularization
Laplacian Growth reaches a singularity (cusps) in finite time \cite{84SB}. In
DLA the ultraviolet regularization is provided by the finite size
of the random walkers. However, many researchers believed
\cite{84Pat} that this difference, which for very large
clusters controls only the smallest scales of the fractal patterns,
were not relevant, expecting the two models to lead to the clusters
with the same asymptotic dimensions. While we argue below that
the difference in ultraviolet regularization is indeed not crucial,
we maintain that the two problems are in two different universality
classes.

In this paper we construct a family of growth processes that includes
DLA and a discrete version of Laplacian Growth as extreme members, using the
same ultraviolet regularization.   We thus expose the
essential difference between DLA and Laplacian Growth. DLA is grown serially,
with the field being updated after each particle growth. On the other
hand all boundary points of a Laplacian pattern are advanced in parallel
at once (proportional to $\B.\nabla P$). We show that this
difference is fundamental to the asymptotic dimension, putting the
two problems in different universality classes. An announcement of these
results was presented in \cite{2000BDLP}.

To reach these conclusions we formulate a theory of Laplacian Growth
patterns in terms of iterated conformal maps. Such a theory was successfully
advanced recently for DLA \cite{98HL,99DHOPSS,00DFHP,2001SL}, 
providing for an unprecedented
analytic control of the properties of DLA \cite{00DP,00DLP}. 
By generalizing it to Laplacian
Growth patterns we can enjoy similar advantages, allowing us to address
delicate points that are beyond the scope of direct numerical simulations
and previous analytic attempts.

In Sect. 2 we extend the iterated conformal maps approach to parallel processes of
layer-by-layer growth with varying local growth rates. In Sect. 3 we 
construct a
2-parameter family of parallel growth processes that includes DLA and 
the discrete Laplacian Growth as special (and distinct) cases. 
We demonstrate the relevance of the
two parameters in determining the asymptotic fractal properties of the 
resulting patterns. In Sect. 4 we consider our algorithm for
Laplacian growth and compare it to the exact
dynamics without surface tension. We study the correspondence 
between these models for the early dynamics (before
the appearance of finite time singularities in the latter). In Sect. 5
we offer concluding remarks.
\section{Iterated Conformal Maps for Parallel Growth Processes}

The method of iterated conformal maps for DLA was introduced in \cite{98HL}. 
Here we present a generalization
to parallel growth processes.
We are interested in
$\Phi^{(n)}(w)$ which conformally maps the exterior of the unit
circle $e^{i\theta}$ in the
mathematical $w$--plane onto the complement of the (simply-connected)
cluster of $n$ particles in the physical $z$--plane.
The unit circle is mapped onto the boundary of the cluster. 
In what follows we use the fact that the gradient of the Laplacian field
$\B.\nabla P(z(s))$ is 
\begin{equation}
\B.\nabla P(z(s)) =\frac{1}{{\Phi^{(n)}}^\prime(e^{i\theta})}\ ,\quad
z(s)=\Phi^{(n)}(e^{i\theta})
\ .
\label{gradient}
\end{equation}
Here $s$ is an arc-length parametrization
of the boundary.
The map $\Phi^{(n)}(w)$ is constructed recursively. Suppose that
we have already $\Phi^{(n)}(w)$ which maps to the exterior of a cluster
of $n$ particles in the physical
plane and we want to 
find the map $\Phi^{(n+p)}(w)$ after $p$ additional particles were added to its 
boundary {\em at once}, each
proportional in size to the local value of $|\B.\nabla P|^{\alpha/2}$.
To grow {\em one} such particle we employ the elementary map 
$\phi_{\lambda,\theta}$ which transforms the unit
circle to a circle with a ``bump" of linear size $\sqrt{\lambda}$
around the point $w=e^{i\theta}$. In this paper we employ the elementary map
\cite{98HL}
\begin{eqnarray}
   &&\phi_{\lambda,0}(w) = \sqrt{w} \left\{ \frac{(1+
   \lambda)}{2w}(1+w)\right. \nonumber\\
   &&\left.\times \left [ 1+w+w \left( 1+\frac{1}{w^2} -\frac{2}{w}
\frac{1-\lambda} {1+ \lambda} \right) ^{1/2} \right] -1 \right \} ^{1/2} \\
   &&\phi_{\lambda,\theta} (w) = e^{i \theta} \phi_{\lambda,0}(e^{-i
   \theta}
   w) \,,
   \label{eq-f}
\end{eqnarray}
If we update the field after the addition of this single particle,
then
\begin{equation}
   \Phi^{(n+1)}(w) = \Phi^{(n)}(\phi_{\lambda_{n+1},\theta_{n+1}}(w)) \ ,
   \label{recurs}
\end{equation}
where $ \Phi^{(n)}(e^{i \theta_{n+1}})$ is the point on which the 
$(n+1)$-th particle is grown and 
$\sqrt{\lambda_n}$ is the size of the grown particle 
divided by the Jacobian of the map ${\Phi^{(n)}}^{'}(e^{i \theta_{n+1}})$
at that point.
  
The map $\Phi^{(n)}(w)$ adds on a
new semi-circular bump to the image of the unit circle under
$\Phi^{(n-1)}(w)$. The bumps in the $z$-plane simulate the accreted 
particles in
the physical space formulation of the growth process. For the height
of the bump to be proportional to $|\B.\nabla P(z(s))|^{\alpha/2}$ we need to
choose its area proportional to
$|{\Phi^{(n-1)}}' (e^{i \theta_n})|^{-\alpha}$ (see Eq. (\ref{gradient})),
or
\begin{equation}
\lambda_{n} = \frac{\lambda_0}{|{\Phi^{(n-1)}}' (e^{i \theta_n})|^{\alpha+2}}
\ . \label{lambdanew}
\end{equation}
With $\alpha =0$ these rules produce a DLA cluster.
Next, to grow $p$ (non-overlapping) particles in parallel, we accrete them without
updating the conformal map. In other words, to add a new layer of $p$
particles when the cluster contains $m$ particles, we need to choose $p$
angles on the unit circle $\{\tilde\theta_{m+k}\}_{k=1}^p$. At these
angles we grow bumps which in the physical space are proportional
in size to the gradient of the field around the $m$-particle cluster:
\begin{equation}
   \lambda_{m+k} = \frac{\lambda_0}{|{\Phi^{(m)}}'
   (e^{i \tilde\theta_{m+k}})|^{\alpha+2}} \ , \quad k=1,2\dots, p \ .
\label{layer}
\end{equation} 
After the $p$ particles were added, the conformal map and the field should
be updated. In updating, we will use $p$ compositions of the elementary
map $\phi_{\lambda,\theta} (w)$. Of course, every composition 
effects a reparametrization of the unit circle, which
has to be taken into account. To do this, we
define a series
$\{\theta_{m+k}\}^p_{k=1}$ according to
\begin{equation}
   \Phi^{(m)}(e^{i\tilde\theta_{m+k}})\equiv
   \Phi^{(m+k-1)}(e^{i\theta_{m+k}})\ . \label{tilde}
\end{equation}
Next we define the conformal map used in the next layer growth
according to
\begin{equation}
   \Phi^{(m+p)}(\omega)\equiv
\Phi^{(m)}\circ\phi_{\theta_{m+1},\lambda_{m+1}}\circ
   \dots\circ
   \phi_{\theta_{m+p},\lambda_{m+p}}(\omega) \ . \label{compose}
\end{equation}
In this way we achieve the growth at the images under $\Phi^{(m)}$ of the
points $\{\tilde\theta_{m+k}\}_{k=1}^p$. To compute the $\theta$ series from
a given $\tilde \theta$ series we use Eq.(\ref{compose}) to rewrite
Eq.(\ref{tilde}) in the form
\begin{equation}
e^{i\theta_{m+k}} =\phi^{-1}_{\theta_{m+k-1},\lambda_{m+k-1}}\circ
\dots \circ \phi^{-1}_{\theta_{m+1},\lambda_{m+1}}(e^{i\tilde
  \theta_{m+k}})
\label{connection}
\end{equation}
The inverse map $\phi^{-1}_{\theta,\lambda}$ is given by 
$\phi^{-1}_{\theta,\lambda}(\omega)=
e^{i\theta}\phi^{-1}_{0,\lambda}(e^{-i\theta}\omega)$ with
\begin{equation}
\phi^{-1}_{0,\lambda} = \frac
{\lambda \omega^2  \pm \sqrt{\lambda^2 \omega^4  - 
\omega^2 [1-(1+\lambda)\omega^2][\omega^2-(1+\lambda)]}}
{1-(1+\lambda)\omega^2} \ ,
\label{inverse-map}
\end{equation}
where the positive root is taken for Re $\omega > 0$ and the negative root 
for Re $\omega < 0$. 
We stress that if we
had taken
$\theta_n={\tilde{\theta}}_n$, 
neglecting the effects of reparametrization, we would find 
abnormally small bumps on the tips. The reparametrization tends to move 
arcs that have to be mapped to fjords to regions that are mapped 
to tips. Then bumps that were supposed to grow in fjords where their
size were normal would be pushed to tips where their size becomes
extremely small; small
bumps would appear where they do not belong.

In Fig. 1 we present a schematic diagram of the parallel growth described
above.
\begin{center}
\begin{figure}
\hskip -1.0 cm
\epsfxsize=7truecm
\epsfbox{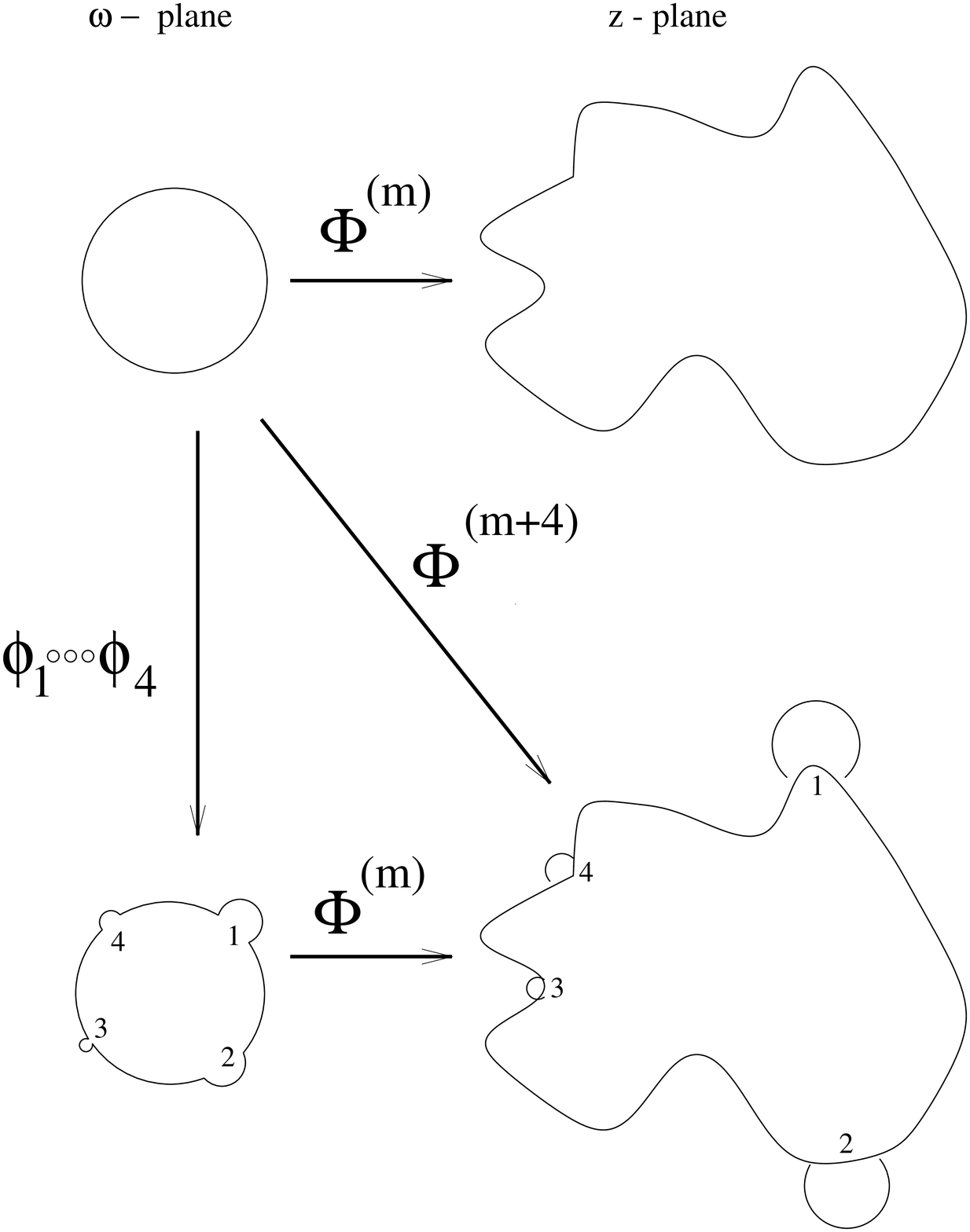}
\caption{A schematic diagram of parallel growth with four particles. The
points 1-4 in the mathematical ($\omega$) plane stand for
${\tilde{\theta}}_1 \dots {\tilde{\theta}}_4$, and are mapped under
$\Phi^{(m)}$ to the appropriate images in the physical ($z$) plane.
The sizes of the bumps were chosen to simulate $\alpha > 0$.}
\label{patterns}
\end{figure}
\end{center}

It is important to notice that on the face of it the conformal map 
(\ref{compose}) appears very
similar to the one obtained in DLA, \cite{98HL,99DHOPSS}. But this is
deceptive. Here we summarize the three major differences between DLA and 
the new growth models:
\begin{itemize}
\item The distribution $\{ \tilde{\theta} \}_{i=1}^{n}$ (which is chosen
uniform in
DLA and related growth models, \cite{00DFHP}) is transformed to a nonuniform
(and maybe even singular) distribution of the angles
$\{ \theta \}_{i=1}^{n}$.
\item The field by which we calculate $\lambda_n$ (i.e. the derivative
of the conformal map) is not updated after each step, but only after
growing a number of particles, and in the limit a whole layer.
\item the parameter $\alpha$ in Eq. (\ref{layer}) can be taken to be
  different than $0$ (which is the value used to grow DLA clusters).
In particular with $\alpha=2$ the size of the bump is proportional
to the gradient of the local field, as is appropriate for Laplacian
Growth. 
\end{itemize}

Note that our algorithm is not purely parallel, as the composition
in Eq.(\ref{compose}) indicates. The parallel aspect is in using
the same {\em field} to compute the values of $\lambda_n$
in Eq. (\ref{layer})
and choosing a uniform distribution of $\{ \tilde{\theta} \}_{i=1}^{n}$
instead of $\{ \theta \}_{i=1}^{n}$. Note that here and below when
we say a ``uniform distribution" we mean that the series was created
without any preference for any region of the unit circle. It very well
may be that after avoiding overlaps the resulting distribution
may be unevenly represented over the unit circle. 
Anyway we make use of an ordered series of compositions of the basic map
$\phi$ to construct one layer.
In the next section we will show
that the serial aspect of the layer growth is not important in terms
of the asymptotic fractal dimension of the clusters. In other
words, the order of placing the bumps is not relevant 
as long as the same field is used as in Eq. (\ref{layer}).

The details of the algorithm, including how to choose
the series $\{\tilde\theta_{m+k}\}_{k=1}^p$ to avoid overlaps
are presented in Appendix A.
\section{two-parameter family of growth processes}

Evidently, a discretized Laplacian Growth 
calls for choosing the series
$\{\tilde\theta_{m+k}\}_{k=1}^p$ such as to have full coverage
of the unit circle (implying the same for the boundary $\Gamma$).
On the other hand DLA calls for growing a single particle 
before updating the field. Since it was shown that in DLA
growth $\lambda_n$ decreases on the average when $n$ increases, in 
the limit of large clusters DLA is consistent with vanishingly small coverage of the
unit circle. To interpolate between these two cases we
introduce a parameter that serves to distinguish one growth model from
the other, giving us a 1-parameter control (the other parameter is $\alpha$). This 
parameter is the {\em degree of coverage}. Since the area covered by the 
pre-image of the $n$-th particle on the unit circle is approximately
$2 \sqrt{\lambda_n}$, we introduce the parameter
\begin{equation}
   {\cal C}=\frac{1}{\pi}\sum_{k=1}^p \sqrt{\lambda_{m+k}} \ .
\end{equation}
(In the Appendix A we show how to measure the coverage exactly).
Since this is the fraction of the unit circle which is covered in
each layer, the limit of Laplacian Growth is obtained with ${\cal C}=1$.
DLA is asymptotically consistent with ${\cal C}=0$. Of course,
the two models differ also in the size of the growing bumps, with
DLA having fixed size particles, ($\alpha=0$ in Eq.(\ref{lambdanew}), 
and Laplacian Growth having particles proportional to $\B.\nabla P$ 
($\alpha=2$ in Eq.(\ref{layer})). 
Together with ${\cal C}$ we have a two parameter control on the
parallel growth dynamics, with DLA and Laplacian Growth 
occupying two corners of the $\alpha, {\cal C}$ plane, at the points
(0,0) and (2,1) respectively.

Needless to say, with our partially serial growth within the layer,
we introduced an extra freedom which is the {\em order} of placement
of the bumps on the unit circle. 
In order for the model to have a physical meaning (i.e. to simulate
true Laplacian growth), it must not depend on the specific itinerary
of $\{ \tilde\theta_i \}_{i=m+1}^{m+p}$ used to cover the unit circle as long as it
is uniformly distributed on the unit circle.
We will show first that this 
extra freedom has no consequence with regards to the asymptotic dimension
of the resulting cluster.
For fixed $\alpha$ the dimension depends only on the value of
${\cal C}$.  To demonstrate this, we will consider various itineraries to
achieve a uniform coverage ${\cal C}$. 
 
One way is to construct the ``golden mean trajectory"
$$\tilde\theta_{m+k+1}=\tilde\theta_{m+k}+2\pi \rho$$
where $\rho=(\sqrt{5}-1)/2$. At each step we check whether the newly
grown bump may overlap a previous one in the layer. If it does, this growth
step is skipped and the orbit continues until a fraction ${\cal C}$
is covered. The first bump of the next layer is
grown at a random position in order to eliminate correlations
induced by the arbitrary itinerary chosen to grow the previous layer.
Another method is random choices of $\tilde\theta_{m+k}$
with the same rule of skipping overlaps. A third method is what was
termed in \cite{00DFHP} the ``period doubling" itinerary
\begin{equation}
\tilde \theta_0=0, \quad \tilde \theta_{2^n+k}=\tilde\theta_k+
\frac{2\pi}{2^{n+1}}\ , \quad 0\le k<2^n \ , n\ge 0
\end{equation}

In all these methods we cannot reach ${\cal C}=1$, since there
are gaps left between the bumps. These are excluded from further
growth in the present layer since their
sizes are smaller than the corresponding value of $\sqrt{\lambda_n}$
as calculated in the middle of the gap.  
We can estimate the maximal value of ${\cal C}$ to be of the order
of 0.65. Nevertheless, to be an acceptable model of parallel
growth the fractal dimension of the resulting cluster should be 
invariant to the itinerary. This invariance is demonstrated below.  
In the next section we treat the case ${\cal C}=1$ by constructing an 
ordered series with extra
care. In the rest of this section we demonstrate the irrelevance 
of the itinerary.

With the present technique it is straightforward
to determine the dimension of the resulting cluster. The conformal map
$\Phi^{(n)}(\omega)$ admits a Laurent expansion
\begin{equation}
   \Phi^{(n)}(\omega) = F_1^{(n)} \omega +F_0^{(n)}
   +\frac{F_{-1}^{(n)}}{\omega}+~\cdots \ .
\end{equation}
The coefficient of the linear term is the Laplace radius, and was
shown to scale like \cite{98HL,99DHOPSS}
\begin{equation}
   F_1^{(n)}\sim S^{1/D} \ ,
\end{equation}
where $S$ is the area of the cluster,
\begin{equation}
S= \sum_{j=1}^n \lambda_j~ |{\Phi^{(j-1)}}' (e^{i\theta_j})|^2 \ .
\end{equation}
Note that for $\alpha=0$ this and equation (\ref{lambdanew}) imply
that $S=n\lambda_0 $. Indeed for $\alpha=0$ this estimate had
been carefully analyzed and substantiated (up to a factor) in \cite{2001SL}.
On the other hand $F_1^{(n)}$ is given analytically by
\begin{equation}
   F_1^{(n)} = \prod_{k=1}^n \sqrt{(1+\lambda_k)} \ ,
\end{equation}
and therefore can be determined very accurately.

In order to achieve comparable growth rates for different
layers we inflated $\lambda_0$ in Eq.(\ref{layer}) according to
$\lambda_0 \to m \lambda_0$ in the layer composed of $p$ particles
$\{m+k\}_{k=1}^p$. The exact form of inflation is not important; we
introduce it simply to oppose the slowing down due to the
decrease of $\langle \lambda_n\rangle$ with $n$. In Fig. 2 we
show
$F_1$ of clusters grown by choosing  the three different itineraries discussed
above to produce the layers and for two
values of ${\cal C}$. We note that the curves superpose for the 3 different clusters
with the same value of ${\cal C}$; for different values
of ${\cal C}$ a different behavior of $F_1$ is manifested.
\begin{figure}
\epsfxsize=7truecm
\epsfbox{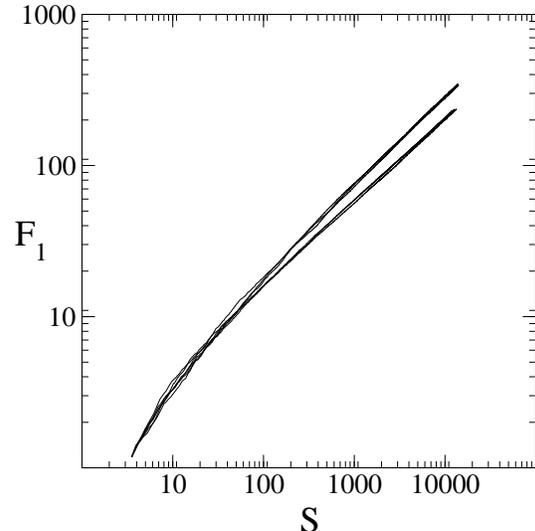}
\caption{Log-log plots of $F_1$ vs. $S$ of six individual clusters with
$\alpha=2$,
using 3 different itineraries for layer construction, with two values of 
${\cal C}$.
${\cal C} =0.3$ (upper group) and ${\cal C} =0.5$ (lower
group). Here we use the golden-mean, random and the period doubling
itineraries (see Ref.[9]).}
\label{Fig2}
\end{figure}
We conclude that the dimension (determined by the asymptotic behavior
of $F_1$ vs. $S$) does not depend on the itinerary used to form the layers
but on ${\cal C}$ only. To understand this further we note that
changing ${\cal C}$ for a fixed value of $\alpha$ is equivalent
to changing the growth probability. For example, if we perform growth
with ${\cal C}=1$ and $\alpha=0$ we generate compact clusters,
since we grow fixed size particles with a uniform measure. ${\cal C}=0$
and $\alpha=0$ is DLA even if we choose a fixed, or even a growing number  of
particles in each layer, as long as ${\cal C}\to 0$ asymptotically.
In Fig.3 we demonstrate this fact by simulating growth with 1 particle
in each layer (DLA), and growth with random addition of particles to
each layer until the first overlap. In both cases  ${\cal C}=0$,
as the number of particles in each layer grows slower than
the available number of sites, and see Sect. 5 for a proof of
this statement. Accordingly the dimension is invariant, despite
the fact that the number of particles in each layer increases
to infinity with the cluster size.
\begin{figure}
\hskip -0.5cm
\begin{center}
\epsfxsize=5truecm
\epsfysize=5truecm
\epsfbox{Fig3a.epsf}
\epsfxsize=8truecm
\epsfysize=8truecm
\epsfbox{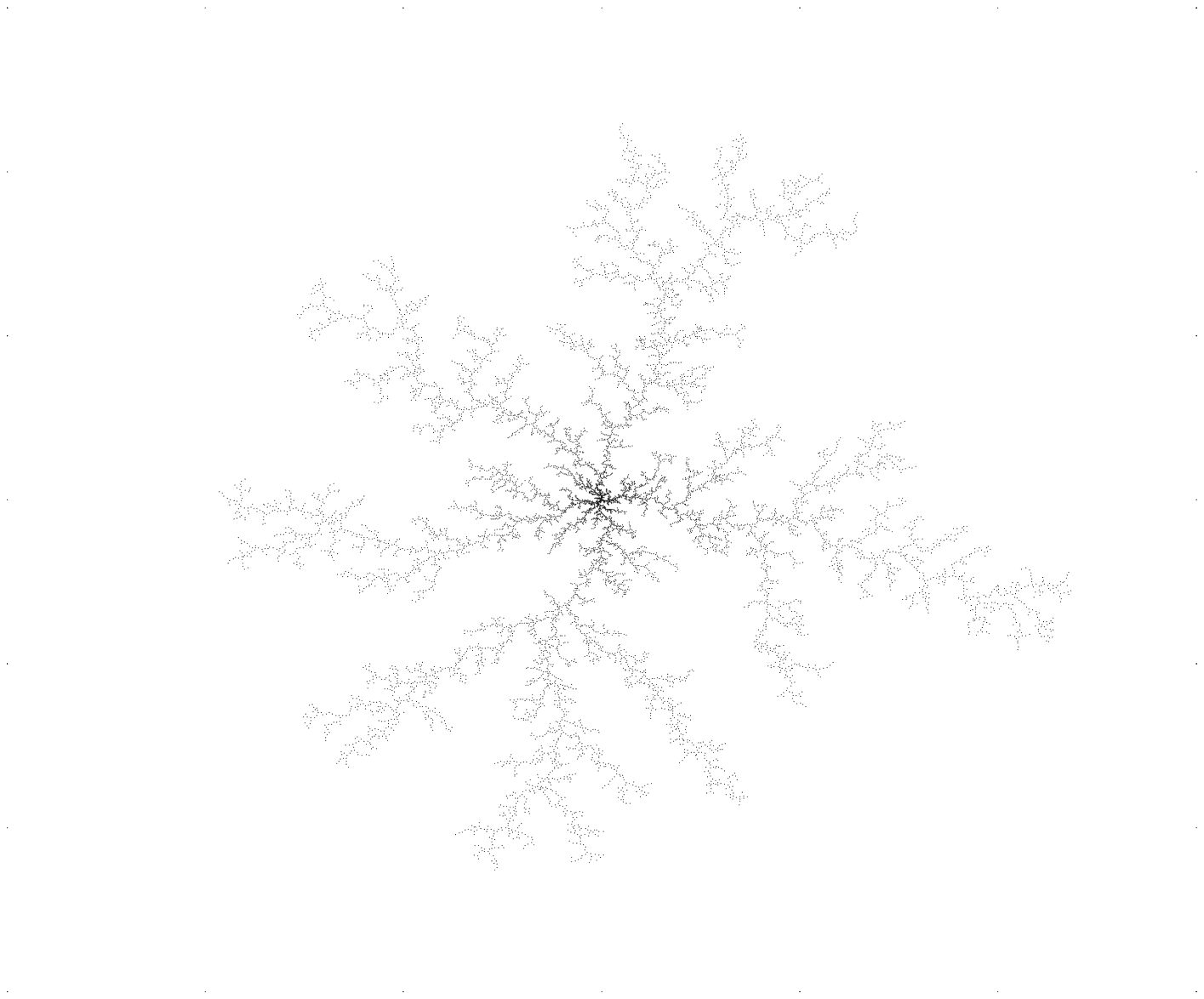}
\caption{(a) Regularly grown DLA (1 particle in a layer). (b) Cluster
grown by randomly attaching bumps in each layer until the first
overlap. Both clusters contain 10 000 particles}
\label{Fig3}
\end{center}
\end{figure}
In Fig.4 we show three fractal patterns grown with 
three different values of ${\cal C}>0$. 
We draw the reader's attention to the fact that drawing cluster like
the one in panel c is not entirely trivial. Simply mapping the unit
circle will not work since many of the fjords will be lost. In fact, 
in appendix \ref{appc} we develop a reliable and effective method to produce
the border of the fractal cluster. 
\begin{figure}
\hskip -0.5cm
\begin{center}
\epsfxsize=5truecm
\epsfysize=5truecm
\epsfbox{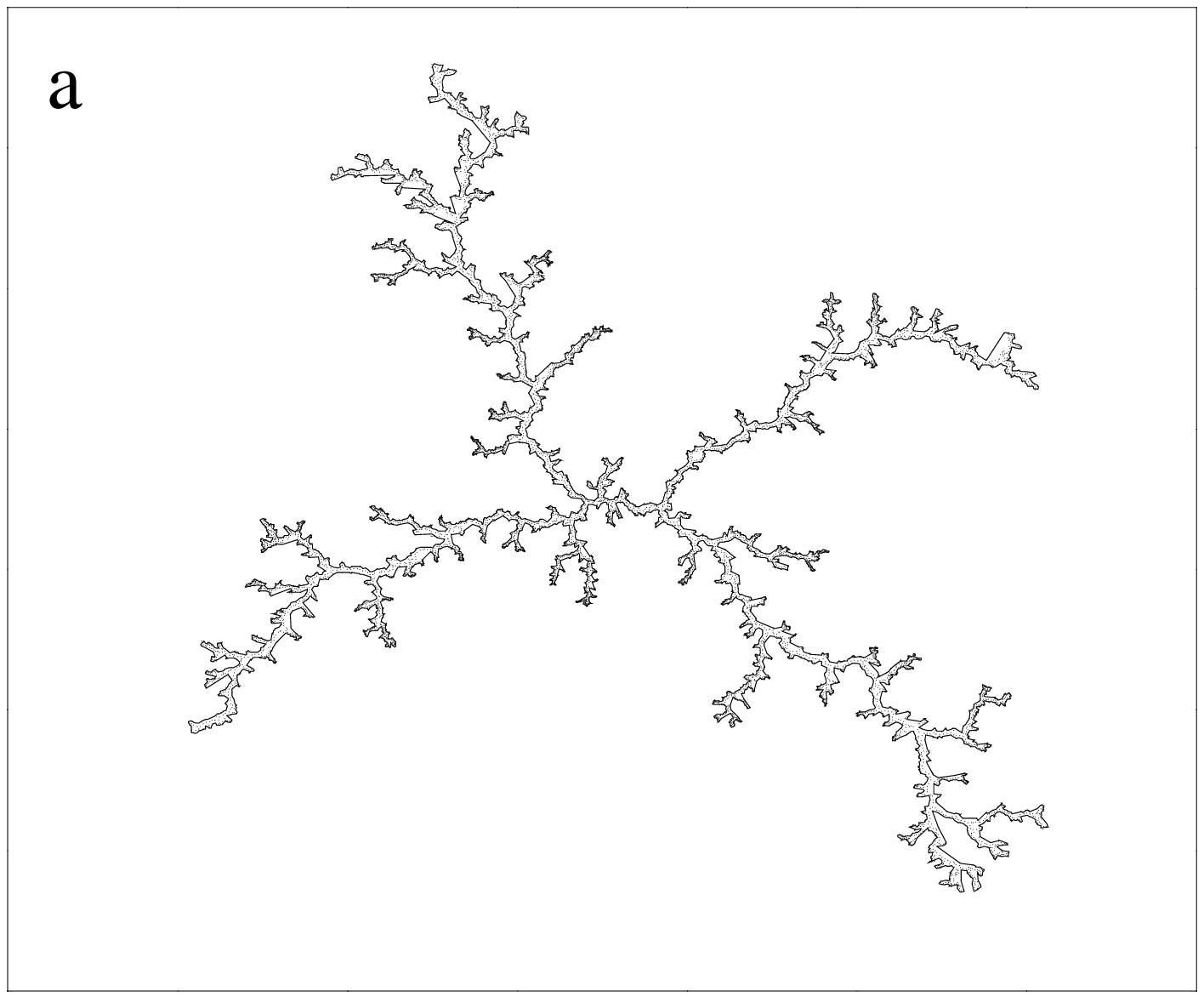}
\epsfxsize=5truecm
\epsfysize=5truecm
\epsfbox{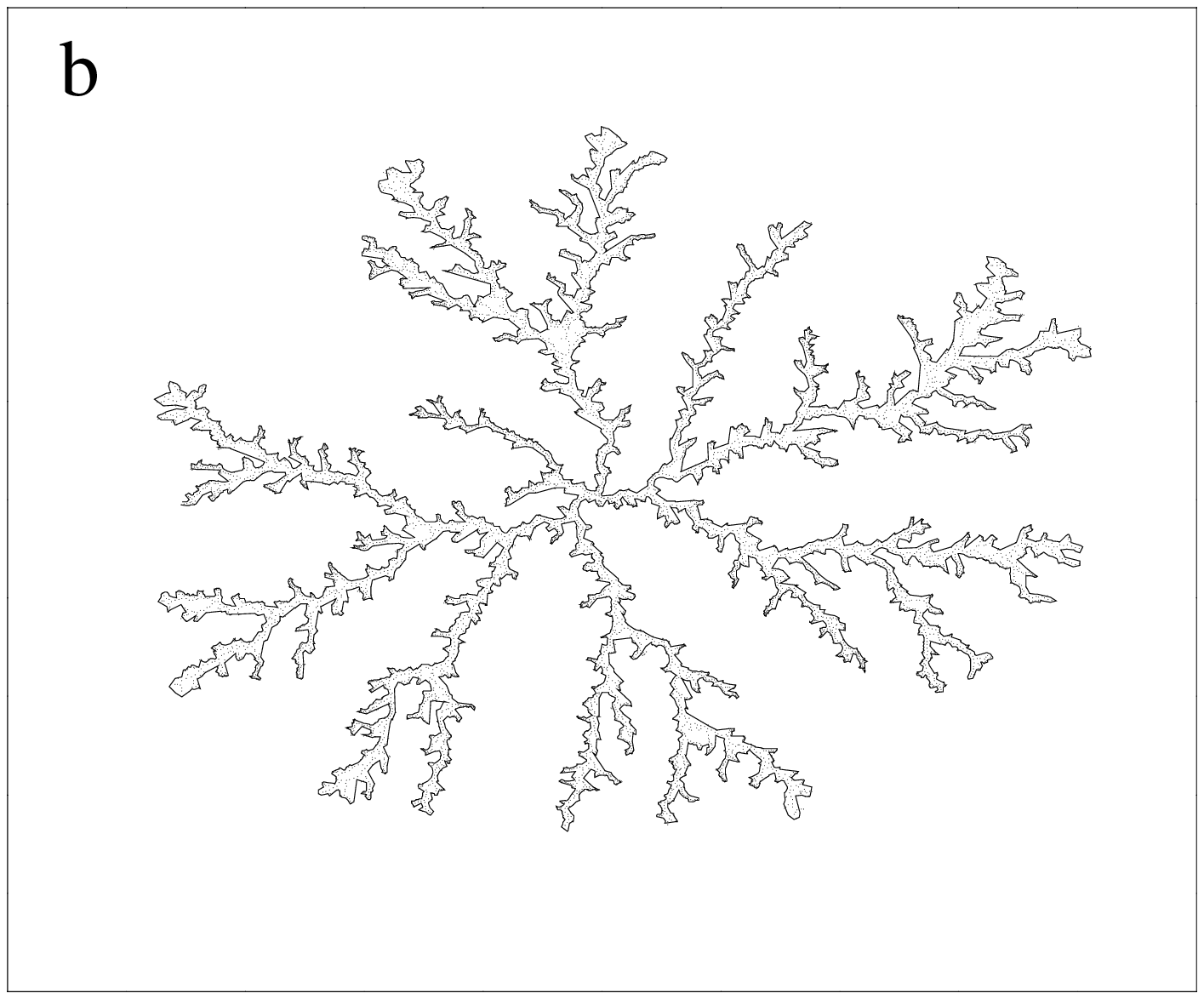}
\epsfxsize=5truecm
\epsfysize=5truecm
\epsfbox{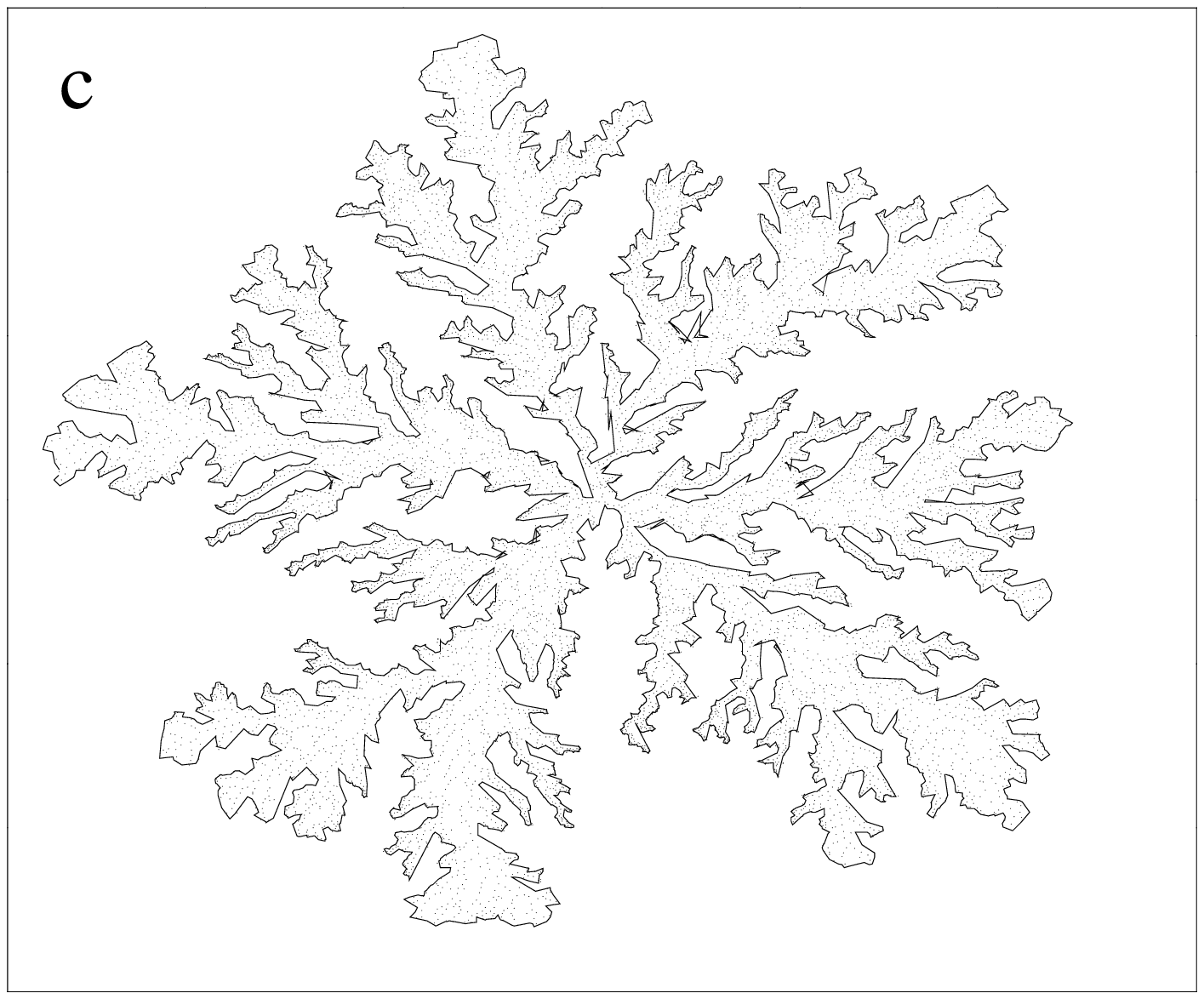}
\caption{Patterns grown with $\alpha=2$ and 3 different values of ${\cal C}$ by using
the golden-mean itinerary: a) ${\cal C} =0.1$, b) ${\cal C} =0.3$,
c) ${\cal C} =0.5$.}
\label{Fig4}
\end{center}
\end{figure}
Even a cursory observation of these patterns
should convince the reader that the dimension of these patterns grows upon
increasing ${\cal C}$.
For a quantitative determination of the dimension we averaged $F_1$ of 
clusters produced by the golden mean itinerary, each with another
random initial angle in each layer.
Plots of the averages $ \langle F_1 \rangle$
for 3 values of ${\cal C}$ are presented in Fig. 5.
\begin{figure}
\epsfxsize=7truecm
\epsfbox{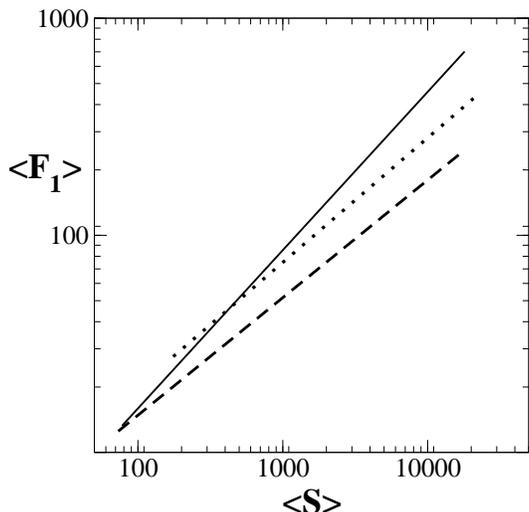}
\caption{Linear regressions of log-log plots of
$ \langle F_1 \rangle$ vs. $S$ for $\alpha=2$ and 3 values of
${\cal C}$: 0.1 (solid line), 0.3 (dotted) and 0.6 (dashed). The slopes
of the curves imply dimensions D=1.37, D=1.75 and D=1.85 respectively.
The averages are taken over 20 clusters.}
\label{Fig5}
\end{figure}
We conclude that the dimension of the growth pattern increases
monotonically with ${\cal C}$, with $D \approx 1.85$ when ${\cal C}=0.6$.

In Fig.6 we present the $\alpha,{\cal C}$
``phase diagram" which results from calculations for a
variety of values of ${\cal C}$ and $\alpha$.
\begin{figure}
\hskip -0.5cm
\epsfxsize=8truecm
\epsfysize=8truecm
\epsfbox{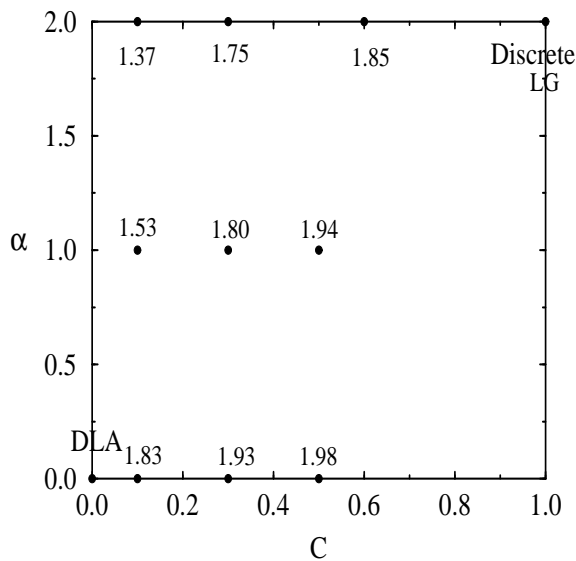}
\caption{``Phase diagram" in which the fractal dimension $D$ is 
displayed for selected values of the parameters ${\cal C}$ and $\alpha$.}
\end{figure}
The conclusion from these
calculations is that the fractal dimension of the clusters
depends continuously on the parameters, growing monotonically
upon decreasing $\alpha$ or increasing ${\cal C}$. It is quite obvious why
increasing ${\cal C}$ should increase the dimension, we simply force particles
into the fjords not allowing them to hit the tips only as is highly probable.
Also decreasing $\alpha$ to $\alpha=0$ increases the dimension, 
since we grow equal
size particles into the fjords, whereas increasing $\alpha$ reduces the
size of particles added to fjords and increases the size of particles
that accrete onto tips. In particular
it is obvious that DLA and our discretized Laplacian Growth cannot have the same
dimensions, putting them in different universality classes.
In particular the dimension $D=1.85$ obtained for $\alpha=2$ and
${\cal C}=0.6$ is a lower bound for the dimension of Laplacian Growth
patterns. This is because the dimension increase with ${\cal C}$
and ${\cal C}=1$ for Laplacian Growth patterns. 
In the next section we present evidence, by reconsidering ${\cal C}=1$, 
that the crucial difference between DLA and Laplacian Growth is not in the
the discretization or in the ultraviolet regularization, but rather 
stems from the different
values of $\alpha$ and ${\cal C}$.

Before turning to models with ${\cal C}=1$ we note in passing that
the present family of models warrants further study on its own
right, independently of the relation between DLA and Laplacian
Growth. The wealth of growth structures seen in electro-deposition,
dielectric breakdown models, and bacterial colony growth \cite{85BGGKLMS} may
very well justify 2-parameter families of models. The present one
is not less physical than any other that had been studied so
far in the literature, but it enjoys the benefit of easily
obtained conformal formulation.
\section{discrete versus continuous Laplacian Growth} 

Continuous Laplacian growth without surface tension
has been studied using dynamics of conformal maps 
in \cite{58ST,84SB}. The dynamical equation for the conformal
map reads
\begin{equation}
Re \{\omega \overline{\Phi'(\omega,t)} \Phi_t(\omega,t)\} =1 \ . \label{SBS}
\end{equation}
As is well known, the solutions of this equation generate
finite time singularities from smooth initial data. 
The simplest example is the initial condition
\begin{equation}
\Phi(\omega,0)=F_1(0) \omega + \frac{F_{-2}(0)}{\omega^2} \ . \label{init}
\end{equation}
The number of Laurent coefficients is preserved by Eq.(\ref{SBS})
with 
\begin{equation}
\frac{F^2_1(t)}{F_{-2}(t)} = {\rm const} \ . \label{const}
\end{equation}
The finite time singularity is seen from the analytic result
(writing $F_j\equiv F_j(0)$)
\begin{eqnarray}
\hskip -1cm
F_1(t) &=& \frac{F_1}{2F_{-2}} \sqrt{ F_1^2 - \sqrt{ [F_1^2
-4F_{-2}^2]^2 - 8tF_{-2}^2}} \nonumber \\
\hskip -1cm
F_{-2}(t) &=& \frac{1}{4 F_{-2}} [ F_1^2 - \sqrt{ [F_1^2 -
  4F_{-2}^2]^2 - 8tF_{-2}^2} \ .
\label{analyticf1f-2}
\end{eqnarray}
At $t\ \to t_c = [F_1^2 - 4F_{-2}^2]^2/8F_{-2}^2$,
$F_1(t)/F_{-2}(t) \to 2$ and a cusp is developed at the images
of $1$, $\exp(\frac{2 \pi i}{3})$, and $\exp(\frac{4 \pi i}{3})$,
see Fig.7 . This simple example motivated attempts to understand
the role of surface tension as an ultraviolet regularization,
see \cite{91DKZ}. We will use this result to study further the
correspondence between our discretized Laplacian growth
and the continuous counterpart,
and to solidify the fundamental difference between the latter
process and DLA.
\begin{figure}
\hskip -0.5cm
\epsfxsize=8truecm
\epsfysize=8truecm
\epsfbox{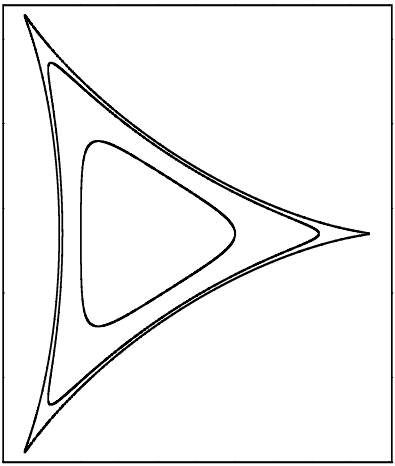}
\caption{Evolution of cusps starting from smooth initial conditions,
with $F_1(0)=1$, $F_{-2}(0)=0.24$. Curves are shown at initial time $t=0$,
an intermediate time and at the critical time $t=t_c$.
}
\end{figure}
As explained in Sec.3, reaching ${\cal C}=1$ is impossible with
any of the itineraries discussed above. We can achieve this
limit by growing in an ordered fashion, adding bumps in
a controlled manner, precisely such as to glue one branch
cut to its neighboring one. How to do this while imposing
the appropriate symmetries is explained in detail in
Appendices A and B. We discover that 
the growth patterns constructed in this way tend to 
fractalize rapidly due to the existence of the branch cuts,
in agreement with our statement above that the result
of our process is a faithful lower bound to the dimension
of continuous Laplacian Growth. An example of the
patterns grown by our discretized process from the initial
conditions (\ref{init}) is shown in Fig.8.
\begin{figure}
\hskip -0.5cm
\epsfxsize=8truecm
\epsfysize=8truecm
\epsfbox{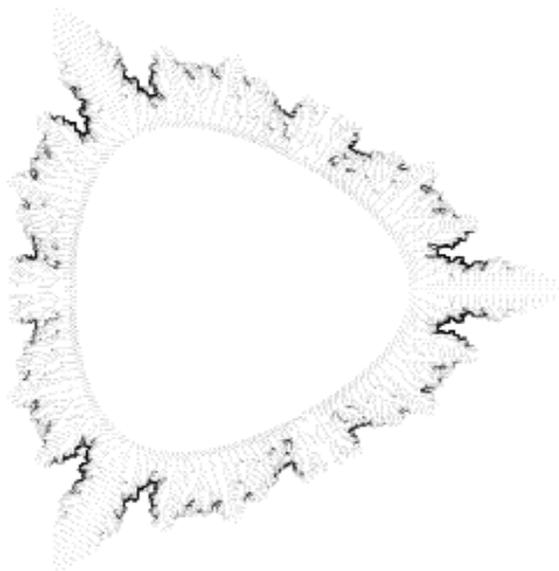}
\caption{Cluster grown with ${\cal C}=1$ starting from the same
initial conditions as in Fig.7. Notice that the branch cuts
lead to spurious fractalization of the smooth envelope.}
\end{figure}
The idea of this section is to isolate the effects of the
parameter ${\cal C}$ and $\alpha$ from effects of discretization
and ultraviolet regularization. To this end we eliminate the 
instabilities caused by the bumpiness
by keeping track of the two Laurent coefficients $F_1$ and $F_{-2}$.
We start with the initial conditions $F_1(0)\omega +F_{-2}(0)/\omega^2$.
Every layer is then  grown by
our algorithm with a chosen values of ${\cal C}$ and $\alpha$, computing 
the new values of $F_1$ and $F_{-2}$, using the analytic
formulae presented in \cite{99DHOPSS}. Discarding all the
other Laurent coefficients we have an updated conformal map
in the form $F^{(n)}_1\omega +F^{(n)}_{-2}/\omega^2$. 

We find the results of this exercise quite revealing. In Fig.9
we show the computed values of $F^{(n)}_1$ and $F^{(n)}_{-2}$ and
the ratio (\ref{const}) for ${\cal C}=1$ and $\alpha=2$,
together with other values of these parameters. 
{\em For ${\cal C}=1$ and $\alpha=2$ the solution approximates
rather closely the exact 
results up to the creation
of the finite time singularity}, with large deviations appearing
only when the tip radius of curvature is of the order of $\lambda_0$.
The degree of approximation improves when $\lambda_0$ is reduced.
On the other hand, the same procedure with other values of 
${\cal C}$ or $\alpha$ deviates from the exact results immediately,
with the degree of deviation being monotonic in the difference
in values of ${\cal C}$ from unity and of $\alpha$ from 2. 
\begin{figure}
\hskip -0.5cm
\epsfxsize=8truecm
\epsfysize=8truecm
\epsfbox{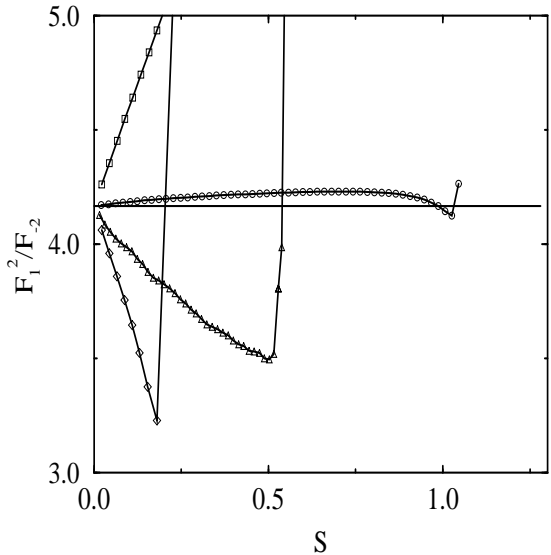}
\caption{$F^2_1/F_{-2}$ as a function of $S$ for the smooth process
described in the text, and for the following values of
${\cal C}$ and $\alpha$: circles, ${\cal C}=1$ and $\alpha=2$;
squares, ${\cal C}=1$ and $\alpha=0$;
diamonds, ${\cal C}=1$ and $\alpha=4$;
triangles, ${\cal C}=0.5$ and $\alpha=2$. The solid line
represents the initial conditions which remain constant, Eq.(19).}
\end{figure}
\begin{figure}
\hskip -0.5cm
\epsfxsize=8truecm
\epsfysize=8truecm
\epsfbox{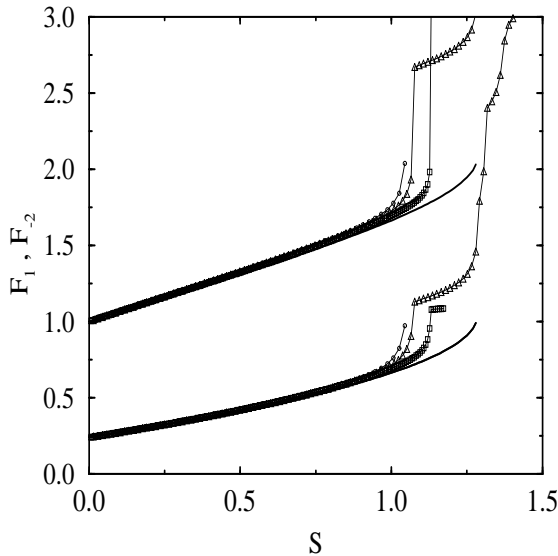}
\caption{$F_1$ and $F_{-2}$ for the smooth growth process described
in the text with ${\cal C}=1$ and three different values of $\lambda_0$.
Circles $\lambda_0=10^{-4}$, triangles $\lambda_0=5*10^{-5}$ and
squares $\lambda_0=10^{-5}$. The solid line results from solving (17)
with the same initial conditions.}
\end{figure}
It is not uninteresting to note the similarity between
the effect of the finite size particles and surface 
tension. This is demonstrated in Fig.10.  The deviation from the 
analytic solution depends on $\lambda_0$. The smaller
the latter is, the deeper we go into the cusp formation,
and the closer we get to the singularity time $t_c$. We estimate
the time of deviation by comparing the radius of curvature
to the physical size of our particle at the tip. This means
that at the tip
\begin{equation}
\frac{\lambda_0}{|\Phi'({\rm tip})|^2} \approx \frac{1}{\kappa^2} \ ,
\end{equation}
where $\kappa$ is the curvature at the tip. The RHS
vanishes when $t\to t_c$, inhibited here by the value of $\lambda_0$.
The time of deviation is therefore when $\lambda_0 = {|\Phi'({\rm
tip})|^2}(t)/\kappa^2(t)$. We can compute the quantities involved analytically:
\begin{eqnarray}
\Phi'({\rm tip}) =F_1-2F_{-2} \ , \\
\kappa= \frac{F_1+4F_{-2}}{(F_1-2F_{-2})^2} \ .
\end{eqnarray}
Accordingly we can estimate the time of deviation and compare
it with the numerics. The agreement is excellent.

At this point it is worthwhile to reexamine the consensus formed
in favor of DLA and Laplacian Growth being in the same universality class.
Superficially one could say that in DLA the update of the harmonic
measure after each particle is not so crucial, since the effect of such an
update is relatively {\em local}
\cite{94Hal}. Thus it may just work that a full layer of particles
would be added to the cluster before major interaction between different growth
events takes place. However this view is completely wrong. An incoming
random walker lands on top of a previously attached one {\em very often}.
To see this, consider how many
angels $\{ \theta_j \}$ can be chosen {\em randomly} on the unit
circle before the first overlap between bumps of linear sizes
$\epsilon_j = \sqrt{\lambda_n (e^{i \theta_j})}$.
To get the order of magnitude take $\epsilon_j =
\epsilon=\langle\sqrt{\lambda_n }\rangle$.
The average number of times that we can choose randomly an angle
before the first overlap
is ${\cal N}(\epsilon
) \sim \frac{1}{\sqrt{\epsilon}}$. The Length
of the unit circle that is covered at that time by the already chosen
bumps is ${\cal L}(\epsilon) = \epsilon {\cal N}(\epsilon) \sim
\sqrt{\epsilon}$.
It was shown in \cite{99DHOPSS} that for DLA $\langle\lambda_n\rangle
\sim \frac{1}{n}$,
so that $\epsilon \sim \frac{1}{\sqrt{n}}$, implying
${\cal N}(n) \sim n^{1/4}$.
Notice that this result means in particular that for a DLA cluster of
1 million particles
only less than 50 random walkers can be attached before two of them
will arrive at the same site! Moreover, ${\cal L}(n) \sim
\frac{1}{n^{1/4}}\to 0$ for $n \to \infty$,
which means that as the DLA cluster grows, our coverage parameter
${\cal C}$
goes to zero, rather than to unity where Laplacian Growth is. Taking
spatial fluctuations of $\lambda_n$ into
account may change the exact exponents but not the qualitative
result.
This argument clarifies the profound difference between growing
a whole layer simultaneously and particle-by-particle. Note however
that DLA is NOT the ${\cal C}\to 0$ limit of our 1-parameter family
because $\alpha = 0$ in Eq. (\ref{lambdanew}) for DLA and $\alpha = 2$ 
in Eq. (\ref{layer}) for Laplacian patterns.
We will now show that if
we eliminate the basic instability that stems from particles landing
on each other {\em then} DLA and Laplacian growth coincide. To do
so we start again with the initial conditions $F_1(0)\omega +F_{-2}(0)/\omega^2$,
grow {\em one particle} with the DLA rules, compute the new value
of $F_1$ and $F_{-2}$, and use the new map $F^{(n)}_1\omega +F^{(n)}_{-2}/\omega^2$
as ``initial conditions" for an additional particle growth. The results
of this process are shown in Fig.11, which is now indistinguishable from
Laplacian growth with ${\cal C}=1$.
In fact, when the instability produced by particles landing one on 
top of the other is eliminated, the choice of the bumps according to
the harmonic measure simulates the growth of a layer, as was expected by 
many researchers. The presence of the instability which is intimately
linked to the DLA growth rules makes it fundamentally different from
the parallel layer growth of Laplacian dynamics.
We believe that with this discussion we offered a conclusive evidence 
for the fundamental difference between DLA and Laplacian Growth.
\begin{figure}
\hskip -0.5cm
\epsfxsize=8truecm
\epsfysize=8truecm
\epsfbox{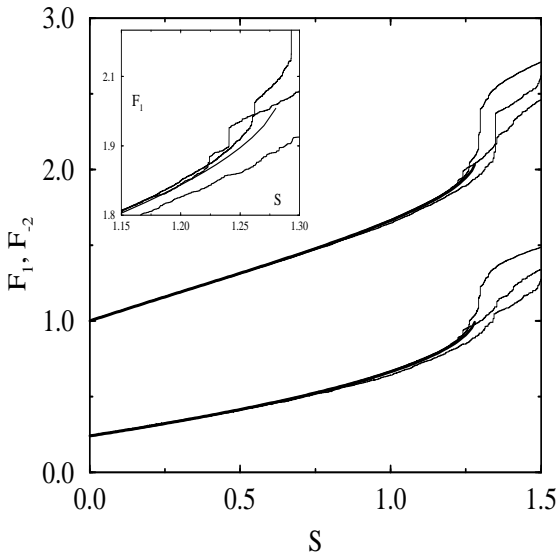}
\caption{Growth Patterns as in Fig.10, starting from the same
initial conditions but growing particle by particle according
to the DLA rules, preserving only $F_1$ and $F_{-2}$. The inset 
shows a zoom of $F_1$ close to the time singularity.}
\end{figure}
\section{Conclusions and Remarks}

We have introduced a 2-parameter family of growth patterns with
the aim of clearly separating DLA from Laplacian Growth. We explained
how to grow in parallel, taking care of the delicate issue of
reparametrization. For the latter issue we needed the inverse map
as explained in Sect. 2. The tools developed to study and control the
reparametrization are further employed to develop symmetry preserving
growth algorithms (Appendix B) and efficient mehtods to construct the
interface of fractal clusters (Apeendix C).
We argued that the parameters ${\cal C}$ and
$\alpha$ are relevant for the asymptotic dynamics, whether the 
order of placing the
bumps is not. The dimensions of the resulting growth patterns were
shown to depend continuously on the two parameters. Besides
providing us with a new model which is interesting by itself,
we could reach the following main conclusions:
\begin{itemize}
\item DLA and Laplacian Growth are not in the same universality
class.
\item The dimension of Laplacian Growth patterns had been bounded
from below by 1.85. We do not have a sharp estimate of this
dimension, and cannot exclude $D=2$.
\item The difference between DLA and Laplacian Growth models
is not in the ultraviolet regularization. We explained that
the deep difference is between the serial and parallel growth
events, leading to increased tendency to form spikes in DLA.
\end{itemize}

In future work it may be worthwhile to attempt to find a sharper
estimate of the dimension of Laplacian Growth patterns. It seems
also worthwhile to study the connection of the present model
to models with noise-reduction, and to further understand 
how to interpret the rich phenomenology of electro-deposition
and bacterial colony growth.

\acknowledgments
It is a pleasure to thank Mitchell J. Feigenbaum, H. George E. Hentschel
and Anders Levermann for invaluable discussions and suggestions.
This work has been supported in part by the
European Commission under the TMR program and the Naftali and Anna
Backenroth-Bronicki Fund for Research in Chaos and Complexity.

\appendix
\section{Details of the Algorithm}
This appendix consists of three parts. In the first we explain
how the absence of overlaps between grown particles 
can be defined in terms of the conformal map. 
The second part of this appendix is dedicated to a detailed description of 
the algorithm that was introduced in Sect. 2. In the last Subsection
we explain the algorithm used to achieve ${\cal C}=1$ at each layer.
\subsection{overlaps in terms of iterated conformal maps}
Suppose that the first particle in a new layer is the $(m+1)$-th particle
in the growth process, and that
there are no overlaps between the first $k$ particles grown in this layer.

In order to express this in terms of the iterated conformal map formalism,
let us make the following definitions:
\begin{itemize}
\item
${\tilde{\omega}}_{n}^{R,L}$ are the two branch points of the map
$\phi_{{\tilde{\theta}}_n,\lambda_n}$, denoted
as ``right" and ``left" respectively.
\item
$\omega_{n}^{R,L}$ are the two branch points of the map 
$\phi_{\theta_n,\lambda_n}$, (which we denote in the 
sequel as $\phi_n$ for brevity).
\end{itemize}
Let us further denote:
\begin{equation}
e^{i \beta_n^{R,L}} \equiv \phi_n (\omega_{n}^{R,L})\ .
\label{beta}
\end{equation}
Note that $|\beta_{n}^{R}-\beta_{n}^{L}|/(2\pi)$ is the fraction of the unit circle
covered by the particle.

The angles $\beta_{n}^{R,L}$ and $\arg [{\tilde{\omega}}_{n}^{R,L}]$ are
connected to each other in a similar manner to the way
${\theta}_n$ is connected to ${\tilde{\theta}}_n$ (\ref{connection}): 
\begin{eqnarray}
e^{i \beta_n^{R}} &=& \phi^{-1}_{n-1}\circ
\dots \circ \phi^{-1}_{m+1}({\tilde{\omega}}_n^R) \nonumber \\
e^{i \beta_n^{L}} &=& \phi^{-1}_{n-1}\circ
\dots \circ \phi^{-1}_{m+1}({\tilde{\omega}}_n^L) \ .
\label{connection-omega}
\end{eqnarray}
Notice that the inverse function  $\phi^{-1}_n$ is analytic on the unit
circle only outside the arc $[\beta_n^R,\beta_n^L]$.

If the $k$-th particle does not overlap any of the previously grown
particles of the layer, then the three points 
$\Phi^{(m+k)}(\omega_{m+k}^R) \ , \ \Phi^{(m+k)}(\omega_{m+k}^L)$, and 
$ \Phi^{(m+k-1)}(e^{i \theta_{m+k}})$ are all in the image of the unit circle
under $\Phi^{(m)}$. In other words, Eqs. (\ref{connection}) and 
(\ref{connection-omega}) (for $n=m+k$) are solvable.

Since $\phi^{-1}_n$ is analytic on the unit circle only outside the 
arc $[\beta_{n}^L, \beta_{n}^R]$, the existence of the following set 
of $k-1$ conditions is necessary and sufficient for the
solvability of Eq. (\ref{connection}).
\begin{eqnarray}
{\tilde{\theta}}_{m+k}
&\notin& [\beta_{m+1}^L,\beta_{m+1}^R] \nonumber \\
\arg [\phi^{-1}_{m+1} ({\tilde{\theta}}_{m+k})] &\notin&
[\beta_{m+2}^L,\beta_{m+2}^R] \nonumber \\
&\vdots& \nonumber \\
\arg [\phi^{-1}_{m+k-2} \circ \dots \circ \phi^{-1}_{m+1} 
({\tilde{\theta}}_{m+k})]&\notin&
[\beta_{m+k-1}^L,\beta_{m+k-1}^R] \ . 
\label{unsolvable}
\end{eqnarray}

Two similar sets of $k-1$ conditions each are obtained by substituting
$\arg [{\tilde{\omega}}_{m+k}^{R,L}]$ instead of ${\tilde{\theta}}_{m+k}$ 
on the LHS of (\ref{unsolvable}), and their existence is 
necessary and sufficient for the solvability of 
Eq. (\ref{connection-omega}).

A failure of any of this conditions means that the 
$k$-th particle is overlapping at least one of the previous $k-1$
particles.
It is clear that if the two edge points of the particle are on the boundary 
of the $m$-particles cluster, so must be its tip (except very rare fill-up 
events \cite{99DHOPSS} that we can safely neglect here). 
Therefore the existence of the last two sets of $k-1$ conditions each is 
sufficient for the existence of the $k-1$ conditions (\ref{unsolvable}). 
\subsection{Growth algorithms for ${\cal C}<1$}
The algorithm for growing one layer of $p$ particles 
on the cluster made up of $m$ particles (given $\Phi^{(m)}$) 
is defined as follows:
\begin{enumerate}
\item
Choose a series 
$\{ {\tilde{\theta}}_{m+k} \}_{k=1}^p$ uniformly distributed on the
interval $[0,2 \pi]$. 

\item 
Define $\theta_{m+1} = {\tilde{\theta}}_{m+1}$.

\item     
Calculate $\lambda_{m+1}$ from Eq. (\ref{layer}), using the 
derivative of $\Phi^{(m)}$ at the point $e^{i {\tilde{\theta}}_{m+1}}$.

\item 
Calculate $\beta_{m+1}^{R,L}$ from Eq. (\ref{beta}) 
and store them.

\item
Let ${\cal C}_1=|\beta_{m+1}^{R}-\beta_{m+1}^{L}|/(2\pi)$
\vskip 0.5cm
For $k>1$:

\item 
Calculate $\lambda_{m+k}$ by the derivative of $\Phi^{(m)}$ at the point
$e^{i {\tilde{\theta}}_{m+k}}$, and find the appropriate branch points
${\tilde{\omega}}_{m+k}^{R,L}$.

\item 
Check the $2(k-1)$ conditions, given in Eq. (\ref{unsolvable}) upon
replacing ${\tilde{\theta}}_{m+k}$ by 
$\arg [{\tilde{\omega}}_{m+k}^{R,L}]$  
on the LHS of Eq. (\ref{unsolvable}).
If any of them is violated (which means that the $k$-th particle overlaps
one of the former $k-1$ particles in the layer), choose another 
${\tilde{\theta}}_{m+k}$ and repeat from stage 6. 

\item
Solve Eqs. (\ref{connection}) and (\ref{connection-omega}) to find
$\theta_{m+k} \ , \beta_{m+k}^R \ , \beta_{m+k}^L$, and store them.

\item
Let ${\cal C}_k={\cal C}_{k-1}+|\beta_{m+k}^{R}-\beta_{m+k}^{L}|/(2\pi)$
\item
After a series of $p$ ``good'' angles $\{ \theta_{m+k} \}_{k=1}^p$ was
found, such that all the $p(p-1)$ solvability conditions 
resulting from $p$ iterations of Eq. (\ref{unsolvable}) are 
fulfilled, update 
the conformal map according to Eq. (\ref{compose}) and ${\cal C}={\cal C}_p$.
\end{enumerate}
In our simulation $p$ is not constant, but is determined by the value of ${\cal C}$
that we want to achieve in each layer.
\subsection{A full coverage (${\cal C}=1$) growth algorithm}
To reach ${\cal C}=1$, we construct recursively a series of consecutive
angles $\{\theta_j\}_{j=m+1}^{m+p}$
such that the left branch cut of the $j$th particle coincides with
the right branch cut of the $(j+1)$th particle. This reads
\begin{equation}
\Phi^{(j+1)}(\omega^R_{j+1})=\Phi^{(j)}(\omega^L_j) \ ,
\end{equation}
or
\begin{equation}
\beta^R_{j+1} = \arg[\phi_{j+1}(\omega^R_{j+1})]= \arg[\omega^L_j] \ .
\end{equation}
Given a pair $(\theta_j, \lambda_j)$ (and hence $\omega^{L,R}_j$ and
$\beta^{L,R}_j$) we have to choose $\theta_{j+1}$ such that the value of
$\beta^R_{j+1}$ which is determined by $\theta_{j+1}$ and the value of
$\lambda_{j+1}$ computed at ${\tilde{\theta}}_{j+1}$ coincides with
the previously computed $\arg[\omega^L_j]$.
Numerically this is obtained as follows.
We start with $\theta_{j+1}$ far enough from $\theta_{j}$. Then, using
Eqs.(\ref{layer}),(\ref{connection}),(\ref{beta}),
we calculate the appropriate values of ${\tilde{\theta}}_{j+1}$,
$\lambda_{j+1}$ and $\beta^R_{j+1}$.
This process is repeated until a value of $\theta_{j+1}$ is found such
that
$0 \le \beta^R_{j+1}-\arg[\omega^L_j] \le 0.01 \sqrt{\lambda_j}$.
We proceed until the whole circle is covered.
\section{Imposing symmetries on the iteration scheme}
In this appendix we explain how to use iterations of conformal maps 
to describe growth in geometries less symmetric than the radial.
In addition we show how to preserve symmetries of the continuous
Laplacian dynamics along the iterations.
The basic idea will be demonstrated through the important example of growth in 
channel geometry, and straightforwardly employed to growth from initial
conditions with reflection symmetry or $n$-fold symmetry in radial geometry.
\subsection{Growth in a channel}
The simplest symmetry that is preserved in the iterations scheme is
$2 \pi$-periodicity. Clearly, $\phi_{\theta,\lambda} (e^{i \zeta}) =  
\phi_{\theta,\lambda} (e^{i (2 \pi + \zeta})$. 
Therefore, if the initial conditions 
(i.e. $\Phi^{(0)}$) have the property 
\begin{equation}
\Phi^{(0)} (e^{i (2 \pi + \theta}) = \Phi^{(0)} (e^{i \theta}) + L 
\label{periodicity-ic}
\end{equation}
where $L$ is the channel width, then $\Phi^{(n)}$ will have this
property for any $n>0$.
The simplest $\Phi^{(0)}$ that has the periodicity property is of course
$\Phi^{(0)}(\omega) = \frac{2}{\pi} \log (\omega)$ ($L=1$) which describes
a growth starting from a flat curve.
Notice that the boundary conditions of the Laplacian field
$\nabla P = \frac{{\Phi^{(n)}}^{'}}{|\Phi^{(n)}|}$ 
at infinity will be automatically changed 
from $\nabla P \sim \frac{\vec{r}}{r^2}$ to 
$\nabla P \sim const \ \ \hat{x}$.  

Suppose now that we want to describe a growth in a channel with 
no-flow boundary conditions at the walls. 
This means that the Laplace problem has to be solved at each
stage with the extra boundary conditions 
that the two walls  $y=0$ and $y=L$ are streamlines of the scalar field $P$
(i.e. $\frac{\partial P}{\partial y}|_{y=0,L}=0$).
Pre-images of streamlines of $P$ in the physical plane are rays 
($\arg(\omega) =const$) in the mathematical plane. Therefore, 
imposing no-flow boundary conditions at the walls amounts to demanding 
that the two rays
$\arg(\omega) = \pm \epsilon \ ( \epsilon \to 0)$ are mapped under 
$\Phi^{(n)}$ to the walls $y=0$ and $y=L$ respectively, for every $n$.

Clearly, the elementary map $\phi_{\theta,\lambda} (\omega)$ does not have
this property. Except for $\theta =0,\pi$ the ray $\arg(\omega) =0$ is
mapped to a curved line in the $z$-plane. Therefore, the appropriate
boundary conditions at the walls are not respected by the iteration process.

We can overcome this difficulty in an analogous way to the image 
method used in electrostatics.
Given initial conditions defined by some ${\tilde{\Phi}}^{(0)}$ we 
construct our $\Phi^{(0)}$ by:
\begin{eqnarray}
\Phi^{(0)}(\omega) &=& {\tilde{\Phi}}^{(0)}(2 \omega)  \ \ \ \ \arg[\omega]
\le \pi \nonumber \\
\Phi^{(0)}(\omega) &=& {\tilde{\Phi}}^{(0)}(2 \pi - \omega) + L \ \ \ \
\arg[\omega] \ge \pi  
\end{eqnarray}
Under $\Phi^{(0)}$ each half of the unit circle is mapped to another copy
of the original interface with reflection symmetry around the real axis 
($\arg [\omega] = 0,\pi$). The pre-image of the two walls $y=0$ and $y=L$ 
under ${\Phi^{(0)}}^{-1}$ are the rays $\arg(\omega) = 0^+$ and  
$\arg(\omega) = \pi$ (or $\arg(\omega) = 0^-$ 
and $\arg(\omega) = \pi$) respectively.   

Now we construct an elementary conformal function that maps
the rays $\arg(\omega) =0,\pi$ onto themselves. This can be achieved by
choosing the elementary map to be 
\begin{equation}
\phi_{\theta,\lambda} \circ \phi_{\bar{\theta},\lambda} (\omega) \ ,
\label{elementary}
\end{equation}
such that the image of the unit circle will have the real line as a symmetry axis.
This is shown schematically in Fig.12. 
\begin{figure}
\hskip -1.0 cm
\epsfxsize=7truecm
\begin{center}
\epsfbox{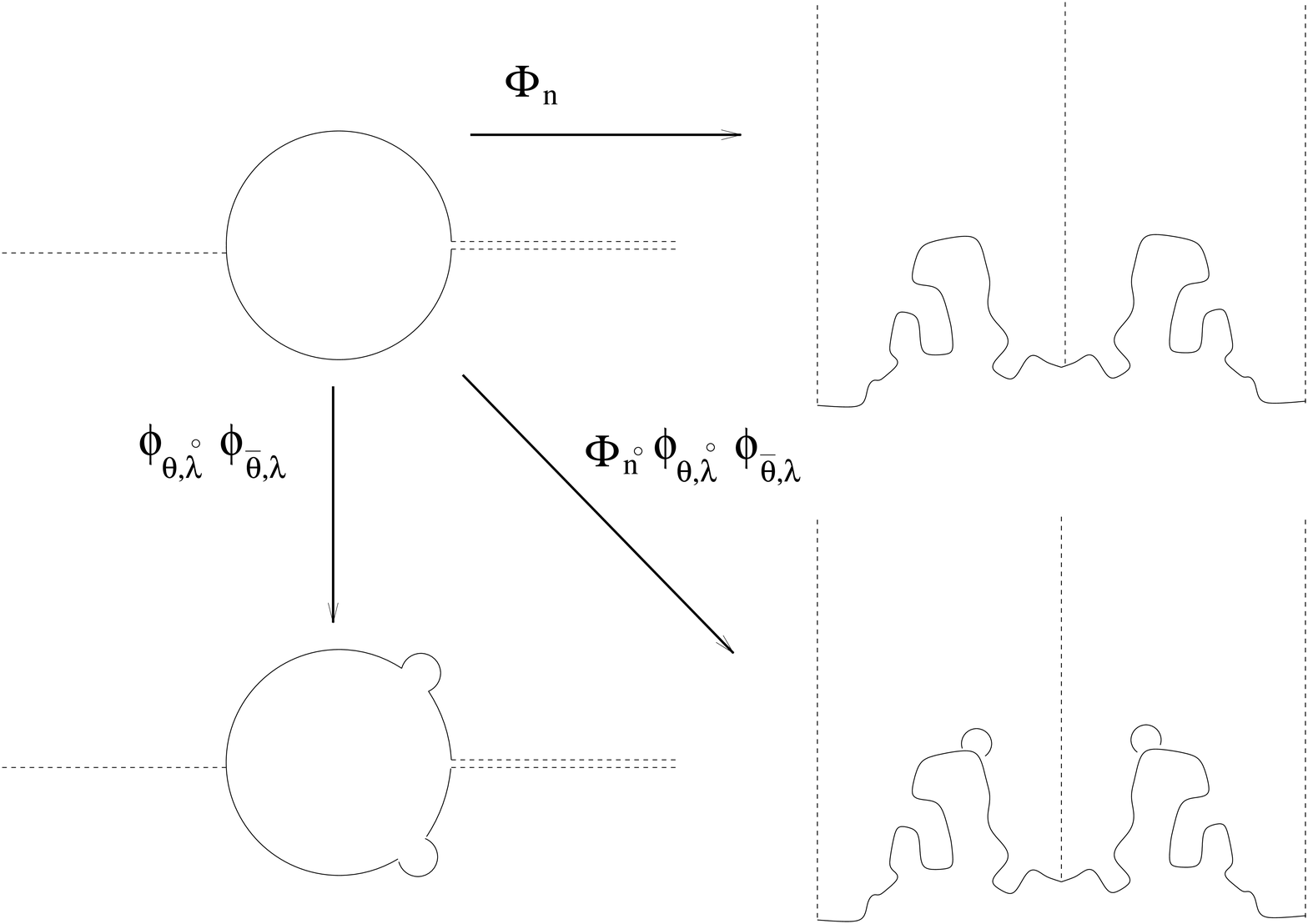}
\end{center}
\caption{Iterative conformal function
that maps at each stage $n$ the unit circle and the real axis in the 
mathematical plane to the evolving interface and
the channel walls in the physical plane, respectively. 
The two rays $\arg [\omega] =0,\pi$ are mapped 
under $\phi_{\theta,\lambda} \circ \phi_{\bar{\theta},\lambda}$ to
themselves, and under the operation of $\Phi^{(0)}$ to the walls $y=0,L$.}
\label{map-channel}
\end{figure}
Since the rays $\arg(\omega) =0,\pi$ are mapped to the walls under 
${\Phi^{(0)}}$ they will be mapped to the walls under ${\Phi^{(n)}}$
defined by
\begin{equation}
{\Phi^{(n)}}(\omega) = {\Phi^{(0)}} \circ 
\phi_{\theta_1,\lambda_1} \circ \phi_{{\bar{\theta}}_1,\lambda_1} \circ \dots 
\circ \phi_{\theta_n,\lambda_n} \circ \phi_{{\bar{\theta}}_n,\lambda_n}
(\omega)
\label{reflective-Phi}
\end{equation} 
Naively, one may think that ${\bar{\theta}} = - \theta$. 
However, constructing the symmetric map as a composition of two
non-symmetric maps leads to some complication.
In order to have a symmetric image of the unit circle, 
one would like to have the second bump in the image of the unit circle to be 
located exactly symmetrically to the first bump:
\begin{equation}
\phi_{\theta,\lambda} (e^{i \bar{\theta}}) = e^{-i \theta} \ .
\label{reparam}
\end{equation}
Eq. (\ref{reparam}) implies choosing $\bar{\theta}$ according to
\begin{equation}
\bar{\theta} = \arg [ \phi^{-1}_{\theta,\lambda} (e^{-i \theta})] \ .
\label{theta-bar} 
\end{equation}
The difference $|\bar{\theta} - (-\theta)|$ becomes smaller with 
$\lambda$, and is zero at the points $\theta = 0,\pi, \pm \pi/2$ for every
value of $\lambda$.
\subsection{Preserving symmetries of the Laplacian dynamics}
The simple technique that was developed in the previous subsection can be
generalized for cases in which a symmetry of the Shraiman-Bensimon
Eq. (\ref{SBS}) is known for specific initial conditions and we want to
preserve it upon using iterations of conformal maps.
\subsubsection{reflection symmetry in radial geometry}
Suppose that the initial interface has a reflection symmetry with respect
to some axis.
Without loss of generality we can take the symmetry axis to be the $x$-axis, 
which is the image under $\Phi^{(0)}$ of the real axis in the mathematical plane. 
Then:
\begin{equation}
\Phi^{(0)} (\omega^*)  = [\Phi^{(0)} (\omega)]^*
\label{reflectionpro}
\end{equation}
It is easy to prove that this symmetry will be preserved under the
Shraiman-Bensimon dynamics.

In order to respect this symmetry in our iterative scheme we use again the
elementary map (\ref{elementary}) which has reflection symmetry with respect 
to the real axis. Thus, $\Phi^{(n)}(\omega)$, defined by Eq. 
(\ref{reflective-Phi}) with $\Phi^{(0)}$ that has the property 
(\ref{reflectionpro}) 
will preserve reflection symmetry.
\subsubsection{$n$-fold symmetry in radial geometry}
The Shraiman-Bensimon Equations preserve also $n$-fold symmetry. Therefore, 
if the initial interface, defined by $\Phi^{(0)}$ has this symmetry, so
should do $\Phi^{(n)}$.
For simplicity let us consider 3-fold symmetry of the form:
\begin{equation}
\Phi^{(0)} (e^{\frac{2 \pi i}{3} } \omega) = e^{\frac{2 \pi i}{3} } \Phi^{(0)}(\omega)
 \ .
\label{3-foldPhi}
\end{equation} 
In order for this symmetry to be preserved, the elementary map must be
3-fold symmetric as well.
Following the discussion in the first part of this appendix
this can be achieved by choosing the elementary map to be
\begin{equation}
\phi_{\theta,\lambda} \circ \phi_{\bar{\theta},\lambda} 
\phi_{\hat{\theta},\lambda}(\omega) \ ,
\label{elementary3fold}
\end{equation}
where
\begin{eqnarray}
\bar{\theta} &=& \arg [ \phi^{-1}_{\theta,\lambda} 
(e^{\frac{2 \pi i}{3} } \theta)]
\nonumber \\
\hat{\theta} &=& 
\arg [ \phi^{-1}_{\bar{\theta},\lambda} \circ \phi^{-1}_{\theta,\lambda} 
(e^{\frac{4 \pi i}{3} } \theta)] \ .
\label{theta-hat} 
\end{eqnarray}
The evolution equation for $\Phi^{(n)}$ now reads
\begin{eqnarray}
{\Phi^{(n)}}(\omega) &=& {\Phi^{(0)}} \circ 
\phi_{\theta_1,\lambda_1} \circ \phi_{{\bar{\theta}}_1,\lambda_1} \circ
\phi_{{\hat{\theta}}_1,\lambda_1}  \circ \dots \circ \nonumber \\
&{\mbox{}}& \circ \dots 
\circ \phi_{\theta_n,\lambda_n} \circ \phi_{{\bar{\theta}}_n,\lambda_n} \circ
\phi_{{\hat{\theta}}_n,\lambda_n} (\omega) \ .
\label{3fold-Phi}
\end{eqnarray}
The extension to higher symmetries is straightforward.
\section{Constructing an outline from branch points}
\label{appc}
The common method \cite{99DHOPSS} to produce the outline of $n$-particles cluster
constructed by the iterated conformal map technique is to sample the unit
circle at $K$ angles $\{ \theta_k \}_{k=1}^{K}$ and to plot their
images under the map $\{ \Phi^{(n)}(e^{i \theta_k}) \}_{k=1}^{K}$. 
This simple method is problematic since a uniform series $\{ \theta_k \}$
will sample the tips much more than the fjords, and thus in order to have a 
reasonable image of the fjords (which are the major part of the fractal
cluster), a huge number $K \gg n$ has to be used. Since calculation of
each image point $\Phi^{(n)}(e^{i \theta_k})$ calls for $O(n^2)$ operations, 
this turns out to be a very inefficient method.

Here we propose an algorithm of $O(n^2)$ complexity to produce an
exhaustive real-space image of the whole cluster.
The key idea is to focus attention on the edge points of the particles,
which are the images of the branch points of the map $\Phi^{(n)}$ on the
unit circle. Each growing particle adds on two new branch points 
to the evolving map and may remove some old ones due to overlaps (see
discussion in Appendix A). Therefore, the number of ``exposed'' branch
points of $\Phi^{(n)}$ is bounded by $2n$.
Let us denote these points $\{ {\omega_k}^{R,L} \}_{k=1}^{n}$.
An exposed branch point ${\omega_{k}}^{R,L}$ was added to the conformal
map by the $k$-th growing particle, and since this particle was not
overlapped by any of the next $n-k$ particles it remains as a branch point of
the map $\Phi^{(n)}$. Nevertheless, the reparametrization of the unit
circle induced by the following $n-k$ iterations changes the pre-image of
each branch point from ${\omega_{k}}^{R,L}$ to ${\omega_{k,n}}^{R,L}$.
The connection between ${\omega_{k}}^{R,L}$ and ${\omega_{k,n}}^{R,L}$ is
given, similarly to Eq. (\ref{tilde}) by:
\begin{equation}
\Phi^{(k)}(\omega_{k}^{R,L}) = 
\Phi^{(n)}(\omega_{k,n}^{R,L})\ , \label{omegank}
\end{equation}
which can be simplified to 
\begin{equation}
\omega_{k,n}^{R,L} =\phi^{-1}_{\theta_{n-1},\lambda_{n-1}}\circ
\dots \circ \phi^{-1}_{\theta_{k+1},\lambda_{k+1}}(\omega_k^{R,L}) \ .
\label{connectionomega}
\end{equation}
The solvability of Eq. (\ref{connectionomega}) determines whether the
appropriate edge point of the $k$-th particle remains exposed under the
addition of the next $n-k$ particles. 
Checking the solvability conditions and calculating the reparametrized
branch points ${\omega_{k,n}}^{R,L}$ from Eq. (\ref{connectionomega}) is
performed in the same way as in Appendix A, and it consists 
of $O([k-n]^2)$ operations. The total complexity of the algorithm is
therefor $O(n^2)$.


\begin{references}

\bibitem{84Pat}
   L. Paterson, Phys. Rev. Lett. {\bf 52}, 1621 (1984);
   L.M. Sander, Nature {\bf 322}, 789 (1986);
   J. Nittmann and H.E. Stanley, Nature {\bf 321}, 663 (1986);
   H.E. Stanley, in ``Fractals and disordered systems'', A. Bunde and
   S. Havlin (Eds.), Springer-Verlag (1991).
  

   \bibitem{58ST}
   P.G. Saffman and G.I. Taylor, Proc. Roy. Soc. London Series A,
   {\bf 245},312 (1958).

 \bibitem{84SB}
   B. Shraiman and D. Bensimon,
   Phys.Rev. A{\bf 30}, 2840 (1984);
   S.D. Howison, J. Fluid Mech. {\bf 167}, 439 (1986).


 \bibitem{81WS} T.A. Witten and L.M. Sander, Phys. Rev. Lett, {\bf 47},
   1400 (1981).

   \bibitem{86BKT}
   D. Bensimon, L.P. Kadanoff, S. Liang, B.I. Shraiman and C. Tang,
   Rev. Mod. Phys. {\bf 58}, 977 (1986);
   S. Tanveer, Phil. Trans. R. Soc. Lond. {\bf A343}, 155 (1993)
   and references therein.
  
   \bibitem{2000BDLP}
   F. Barra, B. Davidovitch, A. Leverman and I. Procaccia
   \rm{arXiv:cond-mat/0103126}.

 \bibitem{98HL} M.B. Hastings and L.S. Levitov, Physica D {\bf 116},
244 (1998).

 \bibitem{99DHOPSS} B. Davidovitch, H.G.E. Hentschel, Z. Olami,
   I.Procaccia,
   L.M. Sander, and E. Somfai,
   Phys. Rev. E, {\bf 59} 1368 (1999).

\bibitem{00DFHP}
   B. Davidovitch, M.J. Feigenbaum, H.G.E. Hentschel and I. Procaccia,
   Phys. Rev. E {\bf 62}, 1706 (2000).
 
\bibitem{2001SL}M. G. Stepanov and L. S. Levitov \rm{arXiv:cond-mat/0005456}
  
\bibitem{00DP}
B. Davidovitch and I. Procaccia, Phys. Rev. Lett., {\bf 85}
       3608-3611 (2000).

\bibitem{00DLP}
B. Davidovitch, A. Levermann, I. Procaccia, Phys. Rev. E, {\bf 62} R5919.

 \bibitem{85BGGKLMS}
   E. Ben-Jacob, R. Godbey, N.D. Goldenfeld, J. Koplik, H. Levine,
   T. Mueller and L.M. Sander, Phys. Rev. Lett. {\bf 55}, 1315 (1985);
   J.D. Chen, Exp. Fluids {\bf 5}, 363 (1987);
   A. Arne\'{o}do, Y. Couder, G. Grasseau, V. Hakim, M. Rabaud,
   Phys. Rev. Lett. {\bf 63}, 984 (1989).

\bibitem{91DKZ}
W-S Dai, L.P. Kadanoff and S-M Zhou, Phys. Rev. A {\bf 43}, 6672 (1991).


 \bibitem{94Hal}
   T.C. Halsey, Phys. Rev. Lett. {\bf 72}, 1228 (1994).

   \end{references}
\end{document}